# Demodulation of Sparse PPM Signals with Low Samples Using Trained RIP Matrix

Seyed Hossein Hosseini, Mahrokh G. Shayesteh, and Mehdi Chehel Amirani

Department of Electrical Engineering, Urmia University, Urmia, Iran

Emails: st_h.hosseini@urmia.ac.ir ; m.shayesteh@urmia.ac.ir ; m.amirani@urmia.ac.ir

**Abstract:** Compressed sensing (CS) theory considers the restricted isometry property (RIP) as a sufficient condition for measurement matrix which guarantees the recovery of any sparse signal from its compressed measurements. The RIP condition also preserves enough information for classification of sparse symbols, even with fewer measurements. In this work, we utilize RIP bound as the cost function for training a simple neural network in order to exploit the near optimal measurements or equivalently near optimal features for classification of a known set of sparse symbols. As an example, we consider demodulation of pulse position modulation (PPM) signals. The results indicate that the proposed method has much better performance than the random measurements and requires less samples than the optimum matched filter demodulator, at the expense of some performance loss. Further, the proposed approach does not need equalizer for multipath channels in contrast to the conventional receiver.

## 1. Introduction

In communication systems, there is a trade-off between the complexity and performance. A challenging complexity, especially at high frequencies, is the number of required samples for demodulation of symbols from the received signal. It can be reduced by extracting some features of the symbol, such as the sparsity. A signal is sparse, if its samples at the Nyquist rate are often zero and rarely nonzero. Compressed sensing (CS) theory makes possible sampling from sparse signals at the information rate, namely less than the Nyquist rate. Compressed sensing has impressed different sciences with its interesting data acquisition framework in various applications, such as seismography, medical imaging, electrocardiography, sensor networks, channel estimation, spectrum sensing, and radar. CS includes two stages: measurement and reconstruction; measurement by a simple matrix multiplication and reconstruction by solving an optimization problem [1-3]. Specializing CS theory for specific applications, in order to achieve better performance, is one of the main challenges in this domain; for example, design of optimum measurement matrix for block sparse signals [4] and modification of original recovery algorithms by using new cost functions for structured sparse signals [5-7]. Meanwhile, there are applications which do not need any reconstruction, such as statistical inference applications, which include detection and classification. In [8-9], the authors presented the principles of compressive classification. An upper bound on the error probability of classification by CS matrices was proposed in [10]. However, since the measurement matrices of CS are general for *any* sparse signal, their classification performance is much less than the optimum matched filter for the given prototype symbols. Nevertheless, it is possible to design a near optimal matrix for compressive classification of a known set of symbols with the minimum number of measurements, even less than that of the matched filter.

Finding optimum measurements for signals which have other features in addition to the sparsity, leads to the performance improvement at the stage of recovery /classification in compressed sensing/classification. For example, multiband signals have block structure in each band, in addition to the sparsity in the Fourier domain (block sparsity). In [4], an algorithm was proposed for finding the optimum measurement of block sparse signals. It is based on the minimization of the inter-block coherence and sub-block coherence between the columns of the recovery matrix. Also, in [11], an appropriate detection matrix has been formulated for sparse signals with known support. It was shown that the performance of the matrix proposed in [11] will be the same as the matched filter, if the number of measurements is equal to the number of non-zero samples of signal.

In this work, we use restricted isometry property (RIP) formulation to find a matrix that satisfies this condition for a known set of sparse symbols. We use a 2-layer neural network for this purpose. After training, the final weights of the first layer of the network will represent the elements of the desired RIP matrix. We show that the

This work was supported by Iran Telecommunication Research Centre (ITRC), Tehran, Iran under contract number 500/9870.

obtained matrix is near optimal for acquisition of sparse symbols which are severely corrupted by noise, from the viewpoints of minimum required measurements along with the reasonable classification error. As an application, we show that the new method achieves a significant reduction in the receiver complexity of pulse position modulation (PPM) signals which can be considered as sparse signals. Sparse PPM signals can be found in TDMA systems with sparse users [12]. Further, our method does not need equalizer in fading channels in contrast to the conventional receiver.

The rest of the paper is organized as follows. In section 2, we briefly review the principles of compressed sensing required for compressive classification. In section 3, we propose a model for design of a measurement matrix that satisfies RIP condition for a known set of sparse symbols. In section 4, we provide simulation results for classification of PPM symbols. Finally, conclusion and suggestions are presented in section 5.

## 2. Compressive Classification

According to the CS theory, any sparse signal, $s \in R^N$ which is specified by a $N$-element vector as its Nyquist samples, can be recovered from its compressively sensed samples, $y \in R^M$, $y = [y_1, y_2, ..., y_M]^T$ by the measurement matrix $\Phi$ of size $M \times N$, that is, $y = \Phi s$ with $M < N$. For this purpose, $\Phi$ must satisfy the following condition for *any* $k$-sparse vector ($k$ is the maximum number of nonzero elements and $k \ll N$):

$$(1-\delta_k)\|s\|_2^2 \leq \|\Phi s\|_2^2 \leq (1+\delta_k)\|s\|_2^2, \quad 0 < \delta_k < 1 \quad (1)$$

where $\|.\|_2$ denotes norm 2. This bound is known as RIP condition for $\Phi$. In [2-3], it was shown that some random matrices such as Gaussian and Bernoulli with $M = O(k \, log \, (N/k))$ (O(.) means order) satisfy the RIP condition by overwhelming probability. However, this value of $M$ is large for classification. In [10], it was demonstrated that random matrices with the number of measurements less than the above bound, has yet enough information for classification with reasonable error probability. In fact, the RIP matrix approximately preserves the angle and Euclidean distance between any pair of $k$-sparse vectors after mapping from $N$-dimensional space to the reduced $M$-dimensional space.

To show the isometric property of $\Phi$, we consider $s_1$ and $s_2$ as the $k/2$ prototype sparse vectors. Then $s_2 - s_1$ will be at most $k$ sparse vector and $\Phi$ under RIP condition for $k$ sparse vectors satisfies:

$$(1-\delta_k)\|s_1 - s_2\|_2^2 \leq \|\Phi(s_1 - s_2)\|_2^2 \leq (1+\delta_k)\|s_1 - s_2\|_2^2 \quad (2)$$

when $\delta_k \to 0$, we have $\|\Phi s_1 - \Phi s_2\|_2^2 = \|s_1 - s_2\|_2^2$, namely, perfect isometry. Fig. 1 demonstrates this mapping. In this figure $s_i$ s, $i = 1, ..., P$ are prototype sparse symbols ($P$ is the number of symbols) and $r$ is the contaminated version an unknown symbol. Although the recovery of $s$ from $y$ and $\Phi$ is possible via different algorithms, but this isometric mapping is only sufficient for classification of a few sparse signals. In other words, we use RIP matrix to extract the fundamental features of sparse vectors for classification.

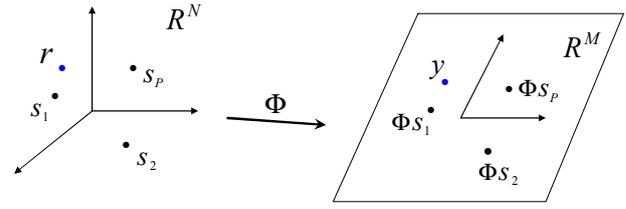

Fig. 1: Isometric mapping of $\Phi$, the distance between $\Phi s_1$ and $\Phi s_2$ approximately is equal to that of $s_1$ and $s_2$.

Fig. 2 illustrates the structure of the compressive receiver for detection of PPM signals. In practice, the $i$-th row elements ($1 \leq i \leq M$) of the matrix $\Phi$ are the samples of the analog waveform $\varphi_i(t)$ at the Nyquist rate of prototype signals $s_i(t)$ (see [11] and references therein). The projection of the received signal into $\varphi_i(t)$ generates the $i$-th element of $y$. The received signal in additive white Gaussian (AWGN) channel is $r(t) = s_i(t) + n(t)$. In this case, the optimum classifier, i.e. maximum likelihood, reduces to the minimum distance classifier. Hence, after obtaining $y$ from the received signal $r$ and measurement matrix $\Phi$, minimum distance of $y$ with the prototype vectors $s_i$ in the reduced space ($M$-dimension) is used for classification. Obviously, for the case of matched filter as a feature selector, we have $\varphi_i(t) = s_i(t)$, $M = P$, and the optimum classifier selects the maximum element of $y$ for detection of received pulse position. Our goal is demodulation with the samples less than the number of symbols ($P$) which is not possible by matched filter.

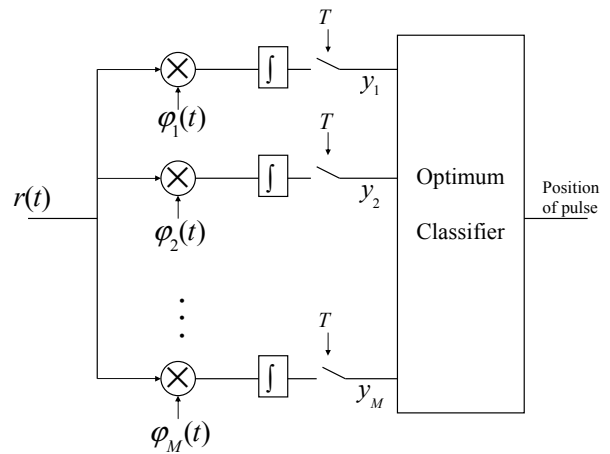

Fig. 2: Block diagram of compressive receiver for demodulation of PPM signals, $T$ is the period of each symbol.

## 3. Proposed Measurement Matrix

Here, we obtain RIP matrix for a specific set of prototype sparse symbols. The RIP condition in Equation (1) can be rewritten as:

$$\left| \|\Phi s\|_2^2 - \|s\|_2^2 \right| \leq \delta_k \|s\|_2^2 \qquad (3)$$

The above notation is considered as the statistical form of RIP. We obtain the RIP matrix of known sparse symbols by training of a two layer neural network. Fig. 3 depicts the neural network model used for this purpose. We aim to preserve the magnitude of the sparse symbols and also the distance between any pair of them, after mapping to the reduced space. Hence, the inputs of neural network are the Nyquist samples or features of the prototype sparse signals/symbols $s_i = [s_{i1},...,s_{iN}]^T$ and the differences between each pair of prototype symbols, $s_i - s_j$, $1 \leq i, j \leq P$; $i \neq j$, totally $L = P + \binom{P}{2}$ inputs shown as $x = \{x_1,...,x_L\}$. The desired output is the squared norm 2 of the input. The weights of the first layer are considered as the elements of the measurement matrix $\Phi$. The outputs of the first layer are the squares of the elements of vector $\Phi x$. The activation function of the first layer is quadratic, i.e. ($f(z) = z^2$) and that of the second layer is linear ($f(z) = z$). The weights of the second layer are set to 1. Hence, the network output is $\|\Phi x\|_2^2$.

The network is trained to minimize the squared error between the net output and the desired output, i.e. $e^2 = (\|\Phi x\|_2^2 - \|x\|_2^2)^2$. A sequential algorithm [13] was used during the learning process; the error propagates through the second layer to the first layer for updating the weights of the first layer. This process repeats until the error becomes less than $\delta_k \|x\|_2^2$ for all learning data. After training, the final weights of the first layer will be the desired RIP matrix.

Updating of the first layer weights is performed using the following recursive algorithm:

$$\Phi^{new} = \Phi^{old} + \mu(\|x\|_2^2 - \|\Phi^{old}x\|_2^2)\Phi^{old}x \times x^T \qquad (4)$$

where $\mu$ is the step size.

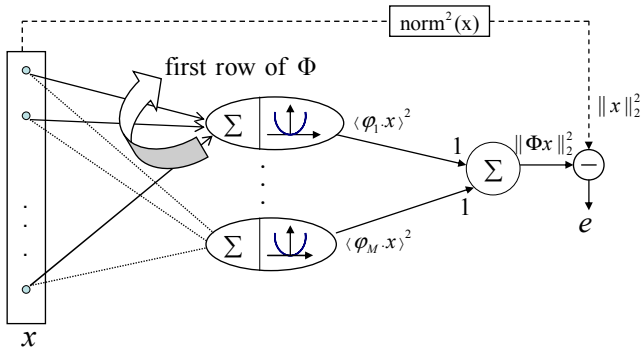

Fig. 3: Neural network model used for training RIP matrix of prototype sparse symbols.

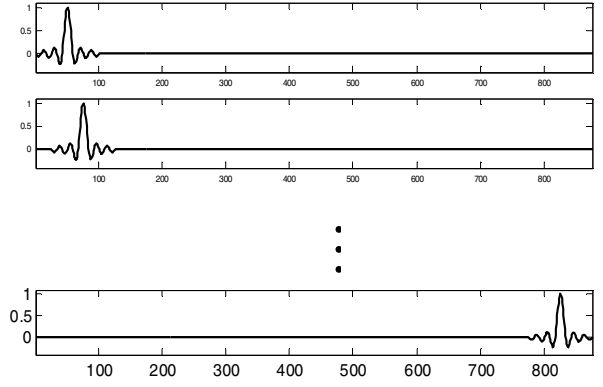

Fig. 4: 32 different PPM Sinc symbols, $N$=876, $k$=100, and the maximum overlap between the adjacent pulses is 75 samples.

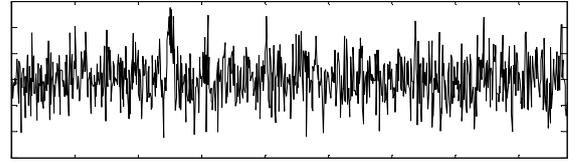

Fig. 5: Corrupted pulse by additive Gaussian noise, $SNR$= -12.3 $dB$, pulse position is in the interval 200-300 (9th position)

## 4. Experimental Results

In order to evaluate the performance of the proposed method, we use 32 PPM symbols, as shown in Fig. 4. The number of Nyquist samples in one symbol period is $N = 876$, the number of nonzero samples is $k = 100$, and the maximum overlap between the adjacent pulses is 75 samples. These prototype vectors ($P = 32$) and the differences between any pair of them (totally $32 + \binom{32}{2} = 544$ vectors) were used as the training data of the neural network explained in the previous section to find the optimum measurement matrix (RIP matrix).

After obtaining the measurement matrix $\Phi$ from the neural network, we use it to extract the fundamental information of the received signal, as shown in Fig. 2. The received signal is one of the prototype symbols corrupted by the additive white Gaussian noise. A typical received signal is shown in Fig. 5. The error probabilities were computed for Gaussian random and the proposed matrices. We also obtained the classification error by the matrix whose rows are prototype symbols, that is, matched filter.

We used different sizes ($M$) for measurement matrix. The neural network parameters are selected in a way to achieve the best performance. We used the parameters $\mu = 0.07$ and $\delta_k \|x\|_2 = 0.2$ for 32 measurements, and $\delta_k \|x\|_2 = 3.8$ for 16 measurements. The initial weight matrix $\Phi_0$ is selected as a random Gaussian with variance of 0.1. We trained the network 10 times. The average numbers of epochs for network convergence were 160 and 210 for 16 and 32 measurements, respectively. Then, we used the obtained RIP matrix in each run to compute the error probability of classification of the data corrupted by noise. Next, we took the average of the errors of 10 trials. Our observations indicate that the variance of error probabilities in 10 times experiments is small, which means the obtained RIP matrices in different runs result in rather the same performance.

Fig. 6 demonstrates the classification error for Gaussian and proposed matrices. Further, we have shown the performance of the optimum matched filter receiver. In computing signal to noise ration (SNR), the energy of signal is obtained by summation the squares of nonzero samples ($k$), and the energy of noise ($\|n\|_2^2$) is estimated by $N\sigma^2$ for different values of noise variance ($\sigma^2$). We observe that the proposed RIP matrix outperforms the Gaussian random matrix significantly. For example, the error probability of the proposed method with 16 measurements ($M=16$) is much less than that of random Gaussian matrix with $M=320$ measurements. Further, the performance of the proposed matrix approaches the optimum matched filter with increasing the number of measurements. In general, there is a trade-off between the number of samples and performance in the proposed method. Note that although the received signal is not sparse because of the presence of additive noise, but the proposed measurement matrix achieves good performance.

Moreover, in order to mitigate the effect of multipath channel in demodulation of sparse PPM signals, it is sufficient to add the Toeplitz matrix of the channel response vector as the first layer of the neural network, i.e. as a fixed layer, and then train the network. However, in the matched filter receiver, we have to use equalizer to combat the channel effect which increases the complexity.

## 5. Conclusion and Suggestions

We obtained RIP matrix for prototype sparse PPM symbols by training a two-layer neural network. The final weights of the first layer are considered as the elements of RIP matrix. Then, we have used a compressive receiver for classification of noisy signals. It was shown that the classification error using the proposed obtained matrix is much less than that of the Gaussian random matrix. The proposed measurement matrix achieves high performance for classification of data which are not even sparse because of the presence of additive noise. The new method introduces significant trade-off between the SNR and the number of required samples for demodulation of PPM sparse symbols. Further, it does not need any equalizer in fading channel in contrast to the conventional receiver.

The proposed method can be also extended to the case when the prototype symbols of classification are sparse or even compressible in the other domains such as Fourier or wavelet or any dictionary which leads to the sparse representation of symbols. In this case, it is enough to add one fix layer to the first layer of neural network as the dictionary matrix.

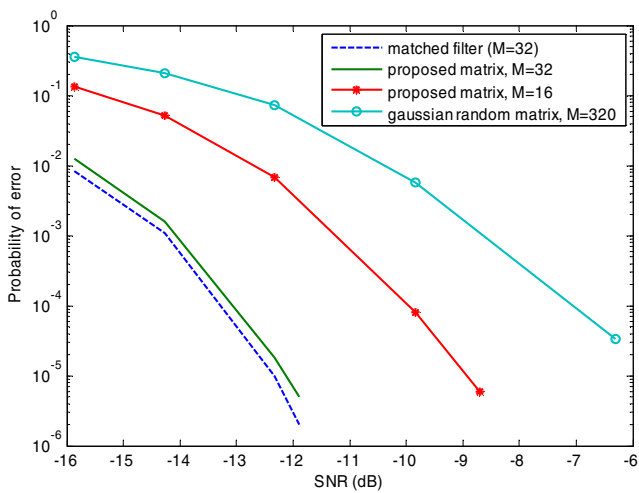

Fig. 6: Probability of error in classification of 32 PPM symbols